 \documentclass[manuscript]{aastex}

 \def\ltsima{$\;\buildrel < \over \sim \;$}
 \def\simlt{\lower.5ex \hbox{\ltsima}}
 \def\gtsima{$\;\buildrel > \over \sim \;$}
 \def\simgt{\lower.5ex \hbox{\gtsima}}
 \begin{document}
 \title{Chemical Composition of the Early Universe}

 \author{Martin Harwit}
 \affil{511 H Street, SW, Washington, DC 20024-2725; also Cornell University}
 \author{Marco Spaans}
 \affil{Kapteyn Astronomical Institute, P.O.\ Box 800, 9700 AV Groningen, The Netherlands}
 \begin{abstract}
 A prediction of standard inflationary cosmology is that the elemental
 composition of the medium out of which the earliest stars and galaxies
 condensed consisted primarily of hydrogen and helium $^4$He with small
 admixtures of deuterium, lithium $^7$Li, and $^3$He.  The most
 red-shifted quasars, galaxies, and Ly-$\alpha$ absorbers currently observed,
 however, all exhibit at least some admixture of heavier elements, as do the
 most ancient stars in the Galaxy.  Here we examine ways in which the
 abundance of these same elements, if present before the epoch of
 population III formation, might be observationally established or ruled out. 
 \end{abstract}
\keywords{Cosmology: Early Universe, Observations, Theory ; Galaxies: Quasars: Absorption Lines}
 \section{Introduction}

 The last few years have seen great advances in the discovery of highly
 red-shifted quasars and galaxies and the study of their spectra.
 High-resolution spectroscopy has simultaneously led to analyses of the
 chemical composition of gases along the line of sight to distant sources.  At
 red shifts of $z\sim 2-4$, Songaila \& Cowie (2001) find a pervasive and
 uniform distribution of ionized carbon and silicon even in relatively tenuous
 intergalactic clouds.  At similar red shifts
 Songaila and Cowie (1996) also found that C\ IV can be detected in 75\% of
 the clouds with column densities $N({\rm HI}) > 3 \times 10^{14}$, while
 Ellison et al.\ (2000) were able to measure C
 IV lines to a limiting sensitivity $\log N({\rm C\ IV}) = 11.75$ against the
 bright lensed quasar Q1422+231. This suggests that the ratio C\ IV / H\ I
 remains roughly constant down to column densities of at least
 $N({\rm HI})\sim 10^{14}$\,cm$^{-2}$ corresponding to roughly the mean
 density of intergalactic gas at the highest red shifts observed $z \sim 3.5$.
 It is not yet clear whether these carbon abundances persist into the
 intergalactic voids.   

 The value of the carbon density parameter that Songaila \& Cowie (2001)
 estimate is of order $\Omega_{{\rm C\ IV}}\sim (1.2\pm
 0.3)\times 10^{-8}\,h^{-1}[1+2q_0z]^{0.5}$.  Taking $h = 0.6$ and a
 deceleration parameter $q_0 =0$ this becomes
 $\Omega_{{\rm C\ IV}}\sim 2\times 10^{-8}$.  And if roughly half the
 carbon is in the triply ionized state, the density parameter for carbon
 integrated over all ionization states should be about
 $\Omega_{{\rm C}}\sim 4\times 10^{-8}$.   For a baryonic density parameter
 corresponding to $\Omega_B \sim 0.04$ this would mean that the carbon
 abundance at these redshifts is a factor
 of 5000 lower than today, i.e., $n_{\rm C}/n_{\rm H} \sim 10^{-7}$.

 Using data on oxygen in various stages of ionization, obtained from
 extreme-ultraviolet absorption lines in Lyman-$\alpha$ absorbers, Telfer
 et al.\ (2002) find an oxygen to carbon ratio [O/C] $\sim 0.3 - 1.2$ at
 comparable red shifts.

 Triply ionized silicon is also observed in the low density clouds.
 The Si\ IV density parameter at $z > 3$ is
 $\Omega_{{\rm Si\ IV}} = 4.3\pm 1.6\times 10^{-9}\,h^{-1}[1+2q_0z]^{0.5}$,
 about a factor of 3 lower than carbon, and with a silicon to carbon ratio
 somewhat higher than in the Sun (Songaila \& Cowie, 2001).

 A number of authors, (Palla, Salpeter \& Stahler, 1983;
 Haiman, Thoul \& Loeb, 1996; Gnedin \& Ostriker, 1997) have suggested that
 a very early generation of population III stars could have formed at red
 shifts ranging out to $z \sim 30$.  Recent simulations due to Abel, Bryan \&
 Norman (2000, 2002) place the epoch of population III formation at
 $z \sim 18$. These massive stars, eventually evolving into type II supernovae,
 would have  produced not only sufficient ionizing radiation to reionize the
 Universe, but would also have produced and explosively ejected heavy elements.
 However, if the population III stars were formed in protogalactic clouds much,
 of the newly formed admixture of heavy elements should have remained
 localized, rather than becoming as dispersed throughout the extragalactic
 medium as the observations of Songaila \& Cowie (2001) indicate.

 The question we now ask is, ``Can observations determine whether a
 significant fraction of the highly dispersed carbon and heavier elements
 could have been part of the primordial mixture from which the first
 generation of stars and galaxies formed or, alternatively, whether all of the
 heavy chemical elements had to be produced in the first rush of star
 formation and then rapidly ejected from galaxies and widely dispersed across
 extragalactic space?" 

 Standard cosmological models argue against the existence of heavy elements as
 admixtures in the primordial brew.  However, an observational confirmation or
 denial of this assumption would provide important insights.  Applegate, Hogan
 \& Scherrer (1988) and  Aguirre (1999) have pointed out some of the
 uncertainties surrounding primordial nucleogenesis, and direct observations
 might suffice to rule out or confirm a number of such possibilities.

 In suggesting observational means for determining the abundances of heavy
 elements {\it before} the formation of population III condensations, care has
 to be taken to determine whether the measurements might not be corrupted by
 detection of heavy elements produced, instead, in the rapidly
 ensuing explosions of these same stars.  Observational tests for determining
 the abundances of such ejecta have recently been proposed by Suginohara,
 Suginohara \& Spergel (1999). If the life of a population III star is
 $\sim 10^6$\,yr at $z \sim 20$, the red shift difference between
 {\it pre-} and {\it post-}collapse will be less than 1\%, making the
 distinction difficult.

 In section 2, we outline potential carbon and oxygen chemistries during the
 dark ages, after recombination but well before population III condensation.
 In section 3, we examine a number of processes that could yield direct
 evidence for primordial carbon or oxygen toward the end of that epoch,
 though all appear too weak for detection with present-day techniques.
 Section 4 suggests a more accessible approach through a search for spectral
 absorption features due to intergalactic neutral or singly ionized carbon or
 neutral oxygen wherever neutral hydrogen Gunn-Peterson troughs are observed.
 A concluding section summarizes these results.

 \section{The Fate of Carbon and Oxygen During the Dark Ages}

 The relative importance of any given chemical reaction, say with rate $R$,
 can be expressed as the cumulative conversion, $\gamma$, per atom,
 ion or molecule integrated over time
 $$\gamma =\int n({\rm H})Rdt=-\int_{z_{\rm min}}^{z_{\rm max}} 2.4\times
 10^{11}R(1+z)^{1/2}dz,$$ with the number density of hydrogen atoms
 $n({\rm H})\sim 4\times 10^{-7}(1+z)^3$ cm$^{-3}$, corresponding to
 a photon-to-baryon ratio $\eta\sim 10^9$ and a current photon density
 $\sim 20T^3\sim 400$ cm$^{-3}$ ($\Omega_B h^2 = 0.034$),
 and $dt\sim -6\times 10^{17}/(1+z)^{5/2}dz$.
 For $z_{\rm max}\sim 1500$ much larger than $z_{\rm min}$, we find
 $\gamma\sim 1\times 10^{16}R$. This cumulative conversion number is given
 for some of the relevant chemical reactions below. The rates of all the
 chemical reactions considered in this work are taken from the UMIST
 database\footnote{http://www.rate99.co.uk/} as described in
 Le Teuff et al.\ (2000).

Because the ionization potentials of oxygen and hydrogen are both close to 13.6 eV, oxygen ions will begin to recombine along with hydrogen at red shifts $z\sim 1500$. As Seager, Sasselov\& Scott (2000) have shown, the ionization fraction of hydrogen depends on the cosmological model chosen, but drops to $\sim 10^{-3}$ at $z < 50$ for currently favored flat cold dark matter models, baryon fractions, and Hubble constants.  Carbon, whose ionization potential is 11.3 eV, will not begin to recombine until $z\sim 1250$, by which epoch the hydrogen ionization fraction will have dropped to $\sim 30 - 50$\%, but will then recombine at temperature-dependent rates 1.5 to 2 times faster.  By $z < 50$ carbon, oxygen and hydrogen ionization fractions should all have dropped to levels of order $10^{-3}$. 

 Neutral carbon atoms suffer
 negligible charge exchange with protons, and hence carbon will, thereafter,
 remain mainly neutral. Chemical processes then are driven mostly by
 neutral-neutral reactions. Furthermore, for modest amounts of H$_2$,
 $n({\rm H_2})/n_{\rm H}<10^{-5}$,
 one expects that the usual initiating reaction
 $${\rm C}^+ + {\rm H}_2\rightarrow {\rm CH}_2^+ + h\nu\ \ \ \ (R = 4.0\times
 10^{-16}[T/300]^{-0.20} {\rm
 cm}^3 {\rm s}^{-1}),\ \gamma\sim 4\ {\rm for}\ {\rm high}\ z,$$
 does not run, nor that
 $${\rm C}^+ + {\rm H}\rightarrow {\rm CH}^+ + h\nu\ \ \ \ (R = 1.7\times 10^{-17} {\rm cm}^3 {\rm s}^{-1}),\
 \gamma\sim 0.17$$
 will be frequent. In any case, CH$^+$ is removed by dissociative recombination
$${\rm CH}^+ + e^-\rightarrow {\rm C} + {\rm H}\ \ \ \ (R = 1.5\times 10^{-7}[T/300]^{-0.42}
{\rm cm}^3 {\rm
s}^{-1}),\ \gamma\sim 10^6\ {\rm if}\ n(e^-)/n_{\rm H}\sim 10^{-4},$$
 although the electron abundance is small.  CH$^+$ will drive a number of rapid reactions with neutral atoms, most notably
$${\rm CH}^+ + {\rm H}\rightarrow {\rm C}^+ + {\rm H}_2\ \ \ \ (R = 7.5\times 10^{-10} {\rm
cm}^3 {\rm s}^{-1}).$$
 The latter reaction then can lead to
 $${\rm C}^+ + {\rm H}_2\rightarrow {\rm CH}_2^+ + h\nu\ \ \ \ (R = 4.0\times
 10^{-16}[T/300]^{-0.20} {\rm
cm}^3 {\rm s}^{-1}),\ \gamma\sim 4n({\rm H}_2)/n_{\rm H},$$
 which depends again on the abundance of H$_2$, and implies that the usual
 ionized carbon chemistry can be driven, albeit at a much lower efficiency
 since both H$_2$ and C$^+$ should have low abundances.

 CH and CH$_2$ formation is initiated through
$${\rm C} + {\rm H}\rightarrow {\rm CH} + h\nu\ \ \ \ (R = 10^{-17} {\rm cm}^3 {\rm
s}^{-1}),\ \gamma\sim 0.1,$$
with
$${\rm CH} + {\rm H}_2\rightarrow {\rm CH}_2 + {\rm H}\ \ \ \ (R = 5.46\times 10^{-10}
e^{-1,943/T} {\rm cm}^3
{\rm s}^{-1}),\ \gamma\sim 10^7n({\rm H}_2)/n_{\rm H},$$
or
$${\rm C} + {\rm H}_2\rightarrow {\rm CH}_2 + h\nu\ \ \ \ (R = 10^{-17} {\rm cm}^3 {\rm
s}^{-1}),\ \gamma\sim 0.1n({\rm H}_2)/n_{\rm H},$$
and
$${\rm C} + {\rm H}_2\rightarrow {\rm CH} + {\rm H}\ \ \ \ (R = 6.64\times 10^{-10}
e^{-11,700/T} {\rm cm}^3 {\rm
s}^{-1},\ \gamma\sim 10^5n({\rm H}_2)/n_{\rm H}\ {\rm at}\ {\rm high}\ z,$$
CH$_2$ is destroyed through
$${\rm CH}_2 + {\rm H}\rightarrow {\rm CH} + {\rm H}_2\ \ \ \ (R = 6.6\times 10^{-11} {\rm
cm}^3 {\rm s}^{-1}),$$
and CH is destroyed through
$${\rm CH} + {\rm H}\rightarrow {\rm C} + {\rm H} + {\rm H}\ \ \ \ (R = 6.0\times 10^{-9}
e^{-40,200/T} {\rm cm}^3 {\rm
s}^{-1}),\ \gamma\sim 2\times 10^3\ at\ z\sim 1500,$$
or
$${\rm CH} + {\rm H}\rightarrow {\rm C} + {\rm H}_2\ \ \ \ (R = 2.7\times
10^{-11}[T/300]^{0.38} {\rm cm}^3 {\rm s}^{-1}),\ \gamma\sim 10^4,$$
which is also a source of H$_2$.

 Note that quite a few of the reactions above have substantial reaction
 barriers or temperature dependencies, so that elevated temperatures are
 required to drive this neutral-neutral chemistry.

 The small abundance of H$_2$ quenches the formation of
 CH$_2^+$ through the otherwise rapid reaction
$${\rm CH}^+ + {\rm H}_2\rightarrow {\rm CH}_2^+ + {\rm H}\ \ \ \ (R = 1.2\times 10^{-9}
{\rm cm}^3 {\rm s}^{-1}),$$
which is normally the dominant removal channel of CH$^+$ in diffuse clouds.
 However, if CH$_2^+$ is present then one has
$${\rm CH}_2^+ + {\rm H}_2\rightarrow {\rm CH}_3^+ + {\rm H}\ \ \ \ (R = 1.60\times
10^{-9} {\rm cm}^3 {\rm s}^{-1}),$$
leading to CH and CH$_2$ after dissociative recombination.  

For oxygen present only in neutral form, and modest amounts of H$_2$,
 $n({\rm H}_2)/n_{\rm H}<10^{-5}$, one expects that the charge transfer
 reaction
$${\rm H}^+ + {\rm O}\rightarrow {\rm O}^+ + {\rm H}\ \ \ \ (R = 7.31\times 10^{-10}
e^{-225.9/T} {\rm cm}^3 {\rm s}^{-1}),$$
which itself is limited by the number of protons, is not followed by
 (the usually rapid) reactions with H$_2$.
 Instead the O$^+$ that is present is removed by the inverse reaction
$${\rm O}^+ + {\rm H}\rightarrow {\rm H}^+ + {\rm O}\ \ \ \ (R = 5.66\times
10^{-10}[T/300]^{0.36} e^{8.60/T} {\rm cm}^3 {\rm s}^{-1}),\ 10-41,000 K,$$
However, one can initiate OH formation through
$${\rm O} + {\rm H}\rightarrow {\rm OH} + h\nu\ \ \ \ (R = 9.9\times 10^{-19}[T/300]^{-0.38}
{\rm cm}^3 {\rm s}^{-1}),\ 10-300 K,$$
followed by reaction with atomic hydrogen that leads mostly back to oxygen in
 atomic form.

 Finally, there are negative ion reactions, similar to the gas-phase formation
 of H$_2$, that lead to OH
$${\rm H} + e^-\rightarrow {\rm H}^- + h\nu\ \ \ \ (R = 4.5\times 10^{-16}[T/300]^{0.6}
e^{4/T} {\rm cm}^3 {\rm s}^{-1}),$$
$${\rm O} + e^-\rightarrow {\rm O}^- + h\nu \ \ \ \ (R = 1.50\times 10^{-15} {\rm cm}^3 {\rm
s}^{-1}),$$
followed by
$${\rm H}^- + {\rm O}\rightarrow {\rm OH} + e^-\ \ \ \ (R = 1.00\times 10^{-9} {\rm cm}^3
{\rm s}^{-1}),$$
$${\rm O}^- + {\rm H}\rightarrow {\rm OH} +e^-\ \ \ \ (R = 5.00\times 10^{-10} {\rm cm}^3
{\rm s}^{-1}),$$
The H$^-$ is rapidly removed through
$${\rm H}^- + {\rm H}\rightarrow {\rm H}_2 + e^-\ \ \ \ (R = 1.30\times 10^{-9} {\rm cm}^3
{\rm s}^{-1}),$$
and by the prevailing background radiation.

For carbon one has similarly
$${\rm C} + e^-\rightarrow {\rm C}^- + h\nu \ \ \ \ (R = 3.00\times 10^{-15} {\rm cm}^3 {\rm
s}^{-1}),$$
followed by
$${\rm H}^- + {\rm C}\rightarrow {\rm CH} +e^-\ \ \ \ (R = 1.00\times 10^{-9} {\rm cm}^3
{\rm s}^{-1}),
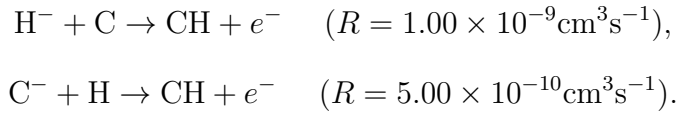
$$
$${\rm C}^- + {\rm H}\rightarrow {\rm CH} +e^-\ \ \ \ (R = 5.00\times 10^{-10} {\rm cm}^3
{\rm s}^{-1}).$$

 This set of chemical pathways is strongly dependent on the
 (generally modest) abundance of free electrons.

 Interactions of carbon with oxygen can be disregarded at the low primordial
 oxygen and carbon abundances, $n_{\rm C,O}/n_{\rm H}\sim 10^{-7}$ permitted
 by direct observations.

 If one evaluates the measure $\gamma$ for the chemical reactions discussed
 above, one finds that both
 carbon and oxygen tend to remain in atomic form. This is largely due to the
 absence of any substantial amount of molecular hydrogen and the long reaction
 time scales relative to the Hubble time.

 \section{Potential Approaches to Directly Detecting Carbon or Oxygen During the Dark Ages}

 We investigated whether the presence of carbon or oxygen would leave an
 imprint on the cosmic microwave background radiation through resonant
 absorption, in analogy to such an effect due to
 primordial traces of lithium (Loeb, 2001).  As pointed out by Loeb (2002),
 however, neither carbon nor oxygen has sufficiently low-lying excited states
 that combine with the ground state
 through allowed transitions. The imprint of both carbon and oxygen through
 this process would be negligibly weak.

 A further weak imprint on the microwave background radiation could arise
 through cooling of gas collapsing to form population III stars.  Both carbon
 and oxygen atoms have fine-structure
 transitions that act as primary coolants in Galactic clouds.  At low
 temperatures they dominate the cooling of molecular hydrogen, whose lowest
 rotational excited level lies at 28.2\,$\mu$m,
 i.e., at an excitation temperature of 510\,K.  This is significantly higher
 than the lowest-lying fine-structure transition of atomic oxygen at
 63.2\,$\mu$m and 228K, or of carbon at 609\,$\mu$m and 23.6\,K.  At the low
 molecular hydrogen concentrations expected at high red shifts and the low
 temperatures expected at the end of the dark ages, cooling through carbon and
 oxygen fine-structure transitions could well prevail over cooling by
 molecular hydrogen through rotational transitions.  The total emitted
 radiation, however, would be low and would fall close to
 the peak of the microwave background radiation.  

 To determine an upper limit to the maximum amount of fine-structure cooling
 to be expected we may consider the collapse of a population III star of mass
 $M$.  The total gravitational energy released in collapsing to a radius
 $r$ is $\sim GM/r$ per unit mass, where $G$ is the gravitational constant.
 The maximum amount of fine-structure emission will be emitted if the
 collapse takes place in thermal equilibrium, since an adiabatic collapse will
 generate no radiation until a shock is formed, and at that point the
 temperature will generally be sufficiently high for
 H$_2$ emission to dominate.  When $GMm_{\rm H}/kr$, exceeds a
 temperature $T \sim 500\,$K, H$_2$ cooling is likely to dominate over
 fine-structure cooling by a factor of $10^5$ even for an abundance of H$_2$
 of $\sim 10^{-2}$; here $k$ is the Boltzmann constant and $m_{\rm H}$ the
 mass of a hydrogen atom.  Hence, fine-structure cooling is not likely to
 exceed $500k \sim 6.9\times 10^{-14}$ erg per carbon or oxygen atom.
 The number of background photons per atom or molecule, however, is the
 entropy $\eta \sim 10^9$. The average energy per photon is $2.73k(1+z)$ for a
 total energy of $2.73\eta k(1+z) \sim 3.8\times 10^{-7}(1+z)$ erg per
 hydrogen atom or molecule.  If, as Abel, Bryan \& Norman (2002) suggest, 
 the dark ages end at $z\sim 20$, the maximum contribution of fine-structure
 transitions to the background would be of order of 1 part in $10^{15}$.
 Because the epoch $\Delta z$ over which the fine-structure radiation took
 place could be quite long, the normally narrow emission lines would be spread
 over a rather broad spectral range, making them even more difficult to
 observe, particularly since emission in thermal equilibrium would mean that
 much of the radiation would occur at a wavelength close to the peak of the
 microwave background.  Detection of this radiation is well short of currently
 foreseeable means.

 \section{A Less Direct Approach}

 Fan et al.\ (2001) have recently detected what may be the long-sought
 Gunn-Peterson trough (Gunn \& Peterson, 1965) in observations of quasars at
 red shifts $z \sim 6$.  They interpret the observed features as due to a
 contiguous expanse of partially neutral intergalactic hydrogen around an
 epoch $z\sim 6$, i.e., before the intergalactic medium became completely
 re-ionized.  The immediate implication of this observation is that much of
 the intergalactic medium remained at least partially neutral for a limited
 period following the formation of population III stars, and that the
 intergalactic medium did not become fully ionized until $z\sim 5$.
 Ionization of the intergalactic medium is certain to have taken
 place {\it before} appreciable contamination through ejection of
 carbon or oxygen from exploding population III stars.  This is partly because
 the number of ionizing photons exceeds the number of ejected ions by many
 orders of magnitude, partly because the ejection velocities of the material
 would be well below the speed of light and incapable of uniformly spreading
 across far reaches of intergalactic space, and partly because most of the
 ejecta will be injected in an ionized state and will long remain ionized at
 the prevailing low intergalactic densities.  Detection of neutral or singly
 ionized carbon, or neutral oxygen wherever a neutral hydrogen Gunn-Peterson
 trough is detected, would therefore indicate the existence of either of these
 elements as part of the
 primordial mixture. A unique signature of the presence of either of these
 elements would be the appearance of one or more additional absorption features
 at wavelengths longer than the Ly-$\alpha$ Gunn-Peterson trough. Singly
 ionized carbon would be expected because radiation at energies between 11.3
 and 13.6 eV would quickly penetrate to great distances across the
 intergalactic medium, ionizing carbon, but leaving hydrogen and oxygen
 neutral. 

 Longward of Ly-$\alpha$, neutral carbon has six particularly strong
 ultraviolet transitions combining with the ground electronic state
 $1s^22s^22p^2 \ ^3P_0$.  For each of these, the ground electronic state is
 split into three fine-structure levels, two of which, respectively lie at
 16.40 and 43.40 cm$^{-1}$ above the true ground state.  The lower of these
 excited fine-structure levels will be significantly populated at red shifts
 above $z \sim 5$, while the upper would have a significant population only
 above $z\sim 15$.  The upper electronic state is also split into three
 levels, leading to six possible fine-structure components to each of the six
 electronic transitions. The six electronic transitions cluster around
 1657, 1561, 1329, 1280, 1277.5,and 1261 \AA.   The products of the oscillator strengths $f$ and statistical weights $g$ for these respective transitions are $fg = 1.26, 0.65, 0.57, 0.19, 0.85$, and 0.36, depending in part on the extent to which the
 lower fine-structure levels are populated.  Longward of Ly-$\alpha$, C$^+$
 has only one significantly allowed transition, a blend of three fine
 structure transitions centered on 1335 \AA, with $fg = 0.76$.  Neutral
 oxygen also has only one strong absorption feature due to a transition at
 1303.5\AA, with $fg = 0.47$. These data and more detailed information on the transitions have been compiled by  Wiese, Fuhr \& Deters (1996).

 Because the cited lines all lie longward of Ly-$\alpha$ they would be
 identifiable even in the presence of strong intergalactic Ly-$\alpha$
 absorption.  However, both carbon and oxygen have a number of relatively strong
 absorption features also shortward of Ly-$\alpha$.  Neutral carbon has a
 range of allowed absorption features at wavelengths between 1194 and
 1189\AA, 1158 and 1156\AA, 1140 and 1138 \AA, and around 1129, 1122 and
 945 \AA.  Singly ionized carbon has just one allowed transition at wavelengths between Ly-$\alpha$ and  the Lyman limit; it lies at 1036.8\AA.  Neutral
 oxygen has allowed transitions around 1040, 1026.5, 989.5,
 977, 972.5, 951.5 \AA, all longward of the Lyman limit at 912\AA.

Following the original work of Gunn and Peterson (1965), the optical depth of these lines is given by 
\begin{equation}
\tau = \left(\frac {\pi e^2 fg}{m_e\nu H_0}\right) n_0 (1+z)^{3/2}
\end{equation}
where $e$ is the electron charge, $m_e$ the electron mass, $\nu$ the spectral frequency of the cluster of transitions, and $H_0$ and $n_0$, respectively the Hubble constant and number density at $z = 0$. Substituting the $fg$ values cited above, we obtain for the neutral carbon 1657\AA\  feature,
\begin{equation}
\tau = 0.21\left(\frac{1+z}{7}\right)^{3/2}\left(\frac {0.6}{h}\right)
\end{equation}
For the five other neutral carbon transitions cited above, we obtain, in order of decreasing wavelengths, respective optical depths of 0.1, 0.08, 0.02, 0.11 and 0.04, for the same red shift and Hubble constant.  For the C$^+$ transition at 1335 \AA, the corresponding optical depth is 0.10 and for neutral oxygen it is 0.06.  At wavelengths between Ly-$\alpha$ and the Lyman limit, the only transitions with appreciable optical depths are the neutral carbon transtion at 945\AA, which has $\tau\sim 0.13$, the C$^+$ transition at 1036.8\AA, with $\tau \sim 0.07$ and the neutral oxygen transitions at 1303 and 989.5\AA, with respective optical dpeths, $\tau \sim 0.06$ and 0.05.

Depending on whether one chooses to search for lines only logward of Ly-$\alpha$, or includes the whole spectral range longward of the Lyman limit, neutral carbon, respectively has four or five absorption features with optical depths $\tau$ in the range of 8 to 20\%; C$^+$ has one or two features with $\tau$ in the range 7 to 10\%, and neutral oxygen one to three features with $\tau \sim 5\%$.

 In the spectrum of the quasar SDSS 1044-0125 at $z\sim 5.73$ obtained by
 Djorgovski et al.\ (2001) there appears to be a narrow trough slightly
 longward of Ly-$\alpha$, at $\sim 8250$\AA\ in the red-shifted spectrum in
 which Ly-$\alpha$ appears to strongly absorb shortward of 8100 \AA.  However,
 none of the longer-wavelength carbon or oxygen lines would lie as close to
 Ly-$\alpha$ as this, since the closest lying line, the 1277.5\AA\ line of
 carbon would lie more nearly at 8500\AA\  for a red-shifted Ly-$\alpha$ line
 at 8100\AA.  Taken at face value, the identification of the Ly-$\alpha$
 trough beginning at 8100\AA\ appears to rule out a significant abundance of
 either primordial atomic carbon or oxygen at levels of order
 $n_{\rm C,O}/n_{\rm H}\sim 10^{-7}$.  This may, however, be because much of
 the intergalactic medium is ionized even at $z = 5.7$ and the broad
 Ly-$\alpha$ trough observed could be due to minuscule traces of neutral
 hydrogen.  For a highly ionized intergalactic medium neutral carbon or oxygen
 features would not lead to sufficient
 optical depth to have been detected, but singly ionized carbon could have
 left a detectable absorption feature longward of the quasar's Ly-$\alpha$
 emission peak.  No such absorption appears to be present in any of the three
 spectra displayed by Fan et al.\ (2001).

 If a more thorough examination shows that absorption features of atomic or ionic carbon or atomic oxygen are consistenly absent in the spectra of distant quasars with strong Gunn-Peterson neutral hydrogen troughs and low levels of ionization, it should be possible to rule out the presence of primordial carbon or oxygen in concentrations four or more orders of
 magnitude lower than solar abundance.

 \section{Discussion}

 To some extent $^{12}$C is the most likely heavy element to exist at early
 times, since unknown circumstances might conceivably contrive to make the
 triple alpha process somewhat efficient at the early epochs of helium,
 deuterium, and lithium formation.  Other elements, like oxygen, would most
 likely have to be formed at far later epochs once stars have formed, and
 would be less likely to exist before the first stars formed.  If carbon
 and/or oxygen did exist before the first stars were formed, the formation of
 heavier elements in the very first generation of massive stars, rather than in
 a second generation, might also be facilitated, which could explain why all
 quasars and galaxies, even those seen at the highest red shifts, already
 exhibit spectra of heavy elements, as do the oldest known Galactic stars.

 The quality of spectra currently available for quasars detected at red shifts of order $z\sim 6$ appear to make possible the detection of absorption features of optical depth $\tau\sim 10\%$ to 20\%, and possibly even weaker features.  With improving techniques, the detection of features with such optical depths should become progressively more reliable.  If the existence of hydrogen Gunn-Peterson troughs {\it before
 substantial re-ionization} can be observationally confirmed, and no additional
 features due to neutral or singly ionized carbon or neutral oxygen absorption
 are found, it should become possible
 to rule out primordial abundances of these two elements at levels as low as
 $10^{-4}$ of the solar abundance and conceivably significantly lower.  

 \section{Acknowledgments}

 The research efforts of M.H.\ are supported by NASA contracts.
 Part of this work was carried out while he was Adriaan Blaauw Visiting
 Professor at the Kapteyn Astronomical Institute in Groningen.  He wishes to
 thank the members of the institute for their kind hospitality, and the Netherlands Organization for Scientific Research (NWO) for their support of the Blaauw Visiting Professorship.
 M.S.\ acknowledges discussions with Andrea Ferrara on the formation of
 pop III stars.

 \section{References}

 Abel, T. R., Bryan, G. \& Norman, M. L. 2000, ApJ, 540, 39

 Abel, T. R., Bryan, G. \& Norman, M. L. 2002, Science, 295, 93

 Aguirre, A. 1999, ApJ 521, 17

 Applegate, J. H., Hogan, C. J. \& Scherrer, R. J. 1988, ApJ 329, 572

 Djorgovski, S. G., Castro, S., Stern, D. \& Mahabal, A. A. 2001, ApJ 560, L5

 Ellison, S. L., Songaila, A., Schaye, J. \& Pettini, M. 2000, ApJ 120, 1175

 Fan, X. et al. 2001, AJ 122, 2833

 Gnedin, N. Y. \& Ostriker, J. P. 1997, ApJ 486, 581

 Gunn J. E. \& Peterson, B. A., 1965, ApJ 142, 1633

 Haiman, Z., Thoul, A. A. \& Loeb, A. 1996, ApJ 464, 523

 Le Teuff, Y. H., Millar, T. J. \& Markwick, A. J., 2000, A\&AS, 146, 157

 Loeb, A. 2001, ApJ 555, L1

Loeb, A. 2002, personal communication to M.H.

 Palla, F., Salpeter, E. E. \& Stahler, S. W. 1983, ApJ 271, 632

Seager, S., Sasselov, D. D. \& Scott, D. ApJS 128, 407

 Songaila, A. \& Cowie, L. L. 1996, AJ 112, 335

 Songaila, A. \& Cowie, L. L. 2001, in {\it The Extragalactic Infrared Background and its
Cosmological Implications}, M. Harwit \& M. G. Hauser, eds., Proceedings of an IAU
Symposium held at Manchester, August 15 - 18, 2000, Astronomical Society of the Pacific.

 Suginohara, M., Suginohara, T., \& Spergel, D. N. 1999, ApJ 512, 547

 Telfer, R. C., et al. 2002, ApJ 579, 500

 Wiese, W. L., Fuhr, J.R. \& Deters, T. M.  1996, {\it Atomic Transition Probabilities of Carbon,
Nitrogen \& Oxygen: a Critical Compilation} Journal of Physical and Chemical Reference
Data, Monograph \#7

\end{document}